\documentclass[aps,twocolumn,prl]{revtex4}
\usepackage{graphicx}

\begin{document}

\preprint{MCTP-05-44}
\preprint{hep-ph/0502177}

\title{Inflating with the QCD Axion}
\date{September 25, 2005}

\author{Katherine Freese}
\email{ktfreese@umich.edu}
\author{James T.~Liu}
\email{jimliu@umich.edu}
\affiliation{Michigan Center for Theoretical Physics,
University of Michigan,
Ann Arbor, MI 48109--1040}

\author{Douglas Spolyar}
\email{dspolyar@physics.ucsc.edu}
\affiliation{Physics Department, University of California,
Santa Cruz, CA 95060}

\begin{abstract} 
  We show that the QCD axion can drive inflation via a series of
tunneling events.  For axion models with a softly broken $Z_N$
symmetry, the axion potential has a series of $N$ local minima and may
be modeled by a tilted cosine.  Chain inflation results along this
tilted cosine: the field tunnels from an initial minimum near the top
of the potential through a series of ever lower minima to the
bottom. This results in sufficient inflation and reheating.  QCD
axions, potentially detectable in current searches, may thus
simultaneously solve problems in particle physics and provide
inflation.
\end{abstract}

\maketitle

In 1981, Guth \cite{guth} proposed an inflationary phase of the
early universe to solve the horizon, flatness, and monopole problems
of the standard cosmology.  During inflation, the Friedmann
equation
\begin{equation}
H^2 = 8\pi G \rho /3 + k/a^2
\end{equation}
is dominated on the right hand side by a (nearly constant) false
vacuum energy term $\rho \simeq \rho_{vac} \sim$ {\it constant}.  The
scale factor of the Universe expands superluminally, $a \sim t^p$ with
$p>1$.  Here $H$ = $\dot a /a$ is the Hubble parameter. With
sufficient inflation, roughly 60 $e$-folds, the cosmological
shortcomings are resolved.

Standard inflationary models require the invention of a new field
whose potential drives the superluminal expansion; there is no direct
evidence that the associated particle exists.  In this paper, we
demonstrate that such a new field is unnecessary: it is possible for
the QCD axion $a$ to drive inflation.  The QCD axion has been proposed
\cite{sw}, \cite{fw} as a solution for the strong CP problem in the
theory of strong interactions.  It is advantageous to use a single
particle to solve several problems.  The mass scales of the axion are
much lower than those of standard inflationary models, and the axion
may be found in ongoing experiments, especially axion searches.

While the axion is {\it a priori} a Goldstone boson of the
spontaneously broken Peccei-Quinn symmetry $U(1)_{PQ}$, QCD instanton
effects induce an axion potential with residual $Z_N$ symmetry.  Our
model includes an additional explicit soft-breaking term, which tilts
the instanton induced potential. While the complete form of the axion
potential is dependent on non-perturbative effects, it is well modeled
as shown in Figure 1 by
\begin{equation}
\label{eq:pot}
V(a) = V_0\left[1-\cos {Na \over v} \right] - \eta 
\cos\left[{a \over v} + \gamma \right] .
\end{equation}
The first term models the periodic instanton potential as a cosine
with $N$ degenerate vacua, or $N$ bumps.  The width of each bump is
given by the Peccei-Quinn scale $f_a=v/N \sim (10^9 - 10^{12}) ~{\rm
GeV},$ and the height of each bump $V_0=m_a^2f_a^2 \sim$ QCD
scale. The second term in (\ref{eq:pot}) is the tilting effect of the
soft-breaking term.  

Inflation starts with the axion field located in
a minimum at the top of the tilted cosine potential.  The universe
tunnels to the next minimum in the cosine, then on down through all
the minima until it reaches the bottom.  The universe inflates a
fraction of an $e$-fold while it is stuck in each of these minima.
Sufficient inflation results for $N \sim$ few hundred.  The general
framework of a sequential chain of tunneling fields was considered
previously in the Chain Inflation model proposed by two of us
\cite{fs}.
\begin{figure}[t]
\centerline{\includegraphics[width=3.0in]{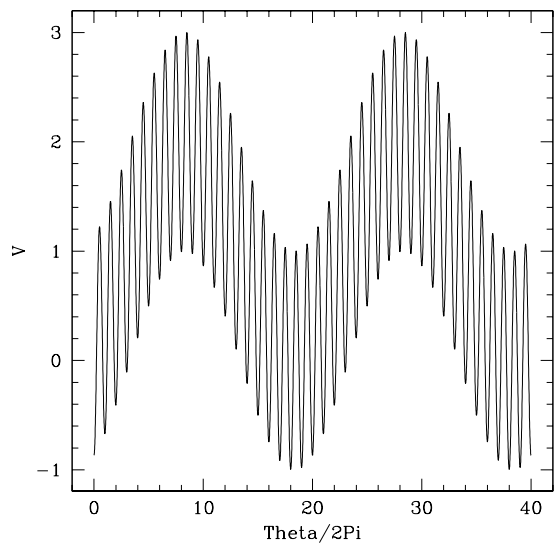}}
\caption{Potential energy density of the QCD axion field $\theta=a/f_a$ as 
a function of $\theta$.  The soft-breaking potential is a tilted cosine as
in Eq.~(\ref{eq:pot}). Here we have taken $N=20$ and $\eta = V_0$.}
\end{figure}

We consider the invisible axion model of Zhitnitskii and Dine,
Fischler, and Srednicki (DFSZ) \cite{z,dfs}.  The axion is identified
as the phase of a complex $SU(2) \times U(1)$ singlet scalar $\sigma$
below the PQ symmetry breaking scale
 
$v/\sqrt{2}$, where $ \sigma = {1 \over \sqrt{2}} (v + \rho)\exp(i
{a \over v}) .  $
 The periodicity of the axion field is $a = a + 2
\pi v
 =a + 2 \pi N f_a$, where $f_a$ is the axion decay constant.
Defining $
 \theta = a/f_a ,
 $ we see that $\theta$ is $2 \pi N$
periodic.  Below the QCD scale $\Lambda_{QCD} \sim 220$MeV, QCD
instantons produce a potential with $N$ degenerate minima at $\theta =
2 \pi n$ where $n=0,1,2,\ldots,N-1$.  The effective action is
\begin{equation}
\mathcal L_{\rm eff} = {1 \over 2} \partial_{\mu} a \partial^{\mu} a
+ (m_a^2f_a^2) g(a/f_a),
\end{equation}
where $g(x)$ is a periodic function of period $2\pi$, Taylor expanded
as $g(x) = g(0) - (1/2) x^2 +\cdots$. The axion mass
is $ m_a \sim 2 N {\sqrt{z} \over 1+z} f_\pi m_\pi /v, $
where $z=m_u/m_d = 0.56$.  Then $m_a f_a \sim m_\pi f_\pi$
where the pion decay constant $f_\pi = 93$MeV and the pion mass
$m_\pi=135$MeV.  Here $N$ refers to the unbroken $Z_N$ subgroup of
$U(1)_{PQ}$, and corresponds to the number and
representations of fermions that carry color charge as well as PQ
charge (and thus contribute to the QCD anomaly) \cite{sikivie}.  Since
we need $N \sim 200$, this introduces additional heavy fermions
beyond the usual quarks and leptons.  The form of $g(x)$ depends
on non-perturbative effects, and hence is not fully specified.  We will
take it to be $g(x)=\cos(x)$.  This captures the main features of the
periodic instanton potential, and will be sufficient for our purposes.

We will take the $U_{PQ}(1)$ symmetry to be softly broken
\cite{sikivie}. Following \cite{sikivie}, we add a soft
breaking term of the form $\mathcal L_{\rm soft} = \mu^3\sigma + h.c.$
where
$\mu$ is a complex parameter and $\sigma$ is given above.  Below the
PQ scale, this adds a term to the axion potential of the form $\eta
\cos(a/v + \gamma)$ where $\eta$ and $\gamma$ are real parameters.
The combined potential is then given in Eq.~(\ref{eq:pot}).  Note that
the phase shift $\gamma$ misaligns the QCD and soft breaking minima.
Without loss of generality, we may restrict $\gamma$ to lie in the
range $-\pi/N < \gamma < \pi/N$.  Away from the bottom of the
potential, the tilt can be treated in the linear regime, so that the
total potential is of the form
\begin{equation}
\label{eq:linear}
V_{\rm linear} = V_0\left[1-\cos {Na \over v} \right] - \eta (a/v+\gamma) .
\end{equation}
The energy difference between minima is roughly $\epsilon \sim
2 \pi \eta /N$.
In this linear regime, the requirement that $\epsilon$ be less
than the barrier height becomes 
$
\epsilon < V_0 .
$
Unless this criterion is satisfied, the barrier becomes irrelevant
and the field simply rolls down the hill.

{\it Sufficient Inflation and Reheating.}
A successful inflationary model has two requirements: sufficient
inflation and reheating.  Here we have a series of a large number
($N$) of tunneling events as the universe transitions from an initial
high vacuum energy down to zero.  In order to have sufficient
inflation, the universe expands by at least 60 $e$-folds once all the
tunneling events have taken place (by the time the field
travels all the way down the cosine):
\begin{equation}
\chi_{tot} > 60 ,
\end{equation}
where $\chi_{tot}$ is the total number of $e$-folds.  In order for the
universe to reheat, the number of $e$-folds attained during {\it one}
tunneling event must be small (less than 1/3 of an $e$-fold), as shown
below.  The failure of old inflation, known as the ``graceful exit''
problem, is circumvented because the bubbles of vacuum are able to
percolate at each step down the potential since the phase transition
is fairly rapid. Chain Inflation's basic mechanism of mulitpile tunneling events \cite{fs}
 works both in the context of the stringy landscape or here with the QCD axion.

In the zero-temperature limit, the nucleation rate $\Gamma$ per unit
volume for producing bubbles of true vacuum in the sea of false vacuum
through quantum tunneling has the form \cite{callan,coleman}
\begin{equation}
\label{eq:tunrate}
\Gamma(t) = A e^{-S_E},
\end{equation}
where $S_E$ is the Euclidean action and where $A$ is a determinantal
factor which is generally the energy scale $\epsilon$ of the phase
transition\footnote{We note that we do not need to include
gravitational effects \cite{deLuccia} as they would only be relevant
for bubbles comparable to the horizon size, whereas the bubbles
considered in this paper are much smaller.}.
 
Guth and Weinberg have shown that the probability of a point remaining
in a false deSitter vacuum is approximately
\begin{equation}
\label{eq:probds}
p(t) \sim \exp({-{4 \pi \over 3}\beta H t}),
\end{equation}
where the dimensionless quantity $\beta$ is defined by
\begin{equation}
\beta \equiv {\Gamma \over H^4 }.
\end{equation}
Writing Eq.~(\ref{eq:probds}) as $p(t) \sim \exp(-t/\tau)$, we
estimate the lifetime of the field in the metastable vacuum as
roughly\footnote{There exists a distribution around this typical
value.}
\begin{equation}
\label{eq:lifetime}
\tau = {3 \over 4 \pi H \beta} = {3 \over 4 \pi} {H^3 \over \Gamma}.
\end{equation}

The number of $e$-foldings for the tunneling event is
\begin{equation}
\label{eq:ni}
\chi = \int H dt \sim H \tau ={3 \over 4 \pi}{H^4\over\Gamma}.
\end{equation}
The authors of \cite{guthwein} and 
\cite{tww} calculated that a critical value of
\begin{equation}
\label{eq:betacrit}
\beta \geq \beta_{crit} = 9/4\pi
\end{equation}
is required to achieve percolation and thermalization.
In terms of $e$-foldings, this is
\begin{equation}
\label{eq:ncrit}
\chi \leq \chi_{crit} = 1/3.
\end{equation} 
As long as this is satisfied, the phase transition at each
stage takes place quickly enough so that `graceful exit' is achieved.
Bubbles of true vacuum nucleate throughout the universe at once, and
are able to percolate.

{\it Tunneling Rate}.
In the thin wall limit, the tunneling rate is given by
Eq.~(\ref{eq:tunrate}).  As shown by \cite{callan,coleman}, we need to
calculate
\begin{equation}
S_1 = \int \sqrt{2 U_+(a)}\, da,
\end{equation}
integrated from one minimum to the next,
where the symmetric portion of the potential is
\begin{equation}
U_+(\theta) = V_0 (1 - \cos \theta) .
\end{equation}
Then
\begin{equation}
S_1 = \sqrt{2 V_0} f_a \int_0^{2 \pi} \sqrt{1-\cos\theta}\,
d\theta = 8 f_a \sqrt{V_0} .
\end{equation}
The Euclidean action is \cite{callan,coleman}
\begin{equation}
\label{eq:euc}
S_0 = {27 \pi^2 S_1^4 \over 2 \epsilon^3}
= 5 \times 10^5 {V_0^2 f_a^4 \over \epsilon^3} .
\end{equation}
For the parameters of the DFSZ axion, $S_0 \gg 1$ and tunneling is
suppressed in the thin wall limit (n.b. the thin wall limit almost never
applies to any realistic tunneling event for any potential as
tunneling is suppressed \cite{weinbergprivatecom}).  To have
reasonably fast tunneling with $\chi <1/3$ from one
minimum to the next, we must be outside the thin wall limit.  
There is the additional constraint $\epsilon < V_0$ in order for
tunneling to take place, as opposed to mere rolling down the
potential.  Thus, obtaining the right amount of inflation requires
$\epsilon/V_0 \sim 1/2$.

{\it The Neutron Electric Dipole Moment.}
We must ensure that the
soft-breaking term in Eq.~(\ref{eq:pot}) does not destroy the strong CP
solution, {\it i.e.}, that the minimum of the potential in Eq.~(\ref{eq:pot})
is not shifted away from zero by more than is allowed by the electric
dipole moment (EDM) of the neutron \cite{Harris:1999jx}
\begin{equation}
\label{eq:EDM}
\Delta \bar \theta \big|_{\rm EDM} < 6\times 10^{-10} .
\end{equation}
To find the minima of the potential, we solve $V'(a)=0$, or
\begin{equation}
V_0 N \sin(N a /v) + \eta \sin(a/v + \gamma) = 0 .
\end{equation}
To leading order in small $\eta$, the minima are located at $a_n =
{2n\pi f_a} - \eta {f_a \over V_0 N }\sin({2\pi n \over N} +
\gamma)$ for integer $n$ where the potential is $V(a_n) = - \eta \cos
({2 \pi n \over N} + \gamma).$ The energy difference
between two adjacent minima is
\begin{equation}
\epsilon = \eta \left[\cos \left({2 \pi n \over N} + \gamma\right)
-\cos \left({2 \pi (n+1) \over N} + \gamma\right) \right ] .
\end{equation}
The difference in field value between minima is
\begin{equation}
\delta a = {2\pi f_a} + \eta {f_a \over V_0 N}
\left[ \sin \left({2\pi n \over N} + \gamma\right) -
\sin\left({2 \pi (n+1) \over N} + \gamma\right) \right] .
\end{equation}

 We have to impose the EDM bounds at the bottom of the
potential, at $n=0$, since this is presumably the endpoint of
tunneling (corresponding to the current universe).  For large $N$, we
find that the shift from $\bar\theta = 0$ is given by
\begin{equation}
\label{eq:thetbottom}
\Delta \bar\theta\big|_{\rm EDM} =  \left|{\eta  \over V_0 N }
\sin\gamma\right|
\sim  \left|{\eta  \over V_0 N }\gamma\right|
\sim {\eta \pi  \over 2 V_0 N^2} .
\end{equation}
In the last equality, we have used that $|\gamma|<\pi/N$ to
estimate that a typical arbitrary value of $\gamma \sim \pi/(2N)$.

During most of the route down the potential, away from the bottom, the
tilt can be approximated as being linear, as in Eq.~(\ref{eq:linear}).
In the linear regime, $\epsilon = 2 \pi \eta /N$.  Taking, {\it e.g.},
$n \sim N/4$, and using Eq.~(\ref{eq:thetbottom}), we find
\begin{equation}
\epsilon_{\rm linear} \sim  4 N V_0 \Delta \bar\theta\big|_{\rm EDM} .
\end{equation}
Combining this with the bound on the neutron EDM, we find that
$
\epsilon_{\rm linear} \leq 2\times 10^{-9} N \Lambda_{QCD}^4,
$
or
$
\epsilon_{\rm linear}^{1/4} \leq 5 (N/200)^{1/4}\hbox{MeV}.
$
To get a sensible reheat temperature $T_R > 10$MeV after inflation%
\footnote{This is the minimal requirement to obtain ordinary element
abundances from nucleosynthesis.  Then baryogenesis must take place
later, as considered by \cite{banksdine1}.}
where $T_R \sim\epsilon_{\rm linear}$,  we need a large number of
new heavy fermions.

At the bottom of the potential, $\epsilon(n=0) \sim 2\pi^2\eta/N^2$, or
using Eq.~(\ref{eq:thetbottom}),
$
\epsilon_{\rm bottom} \sim 4 \pi V_0 \Delta \bar \theta\big|_{\rm EDM} .
$
Combining this with the bound on the neutron EDM, we find that
$
\epsilon_{\rm bottom}^{1/4} \leq 2 {\rm MeV} .
$

The two conditions in the two regimes
%
\begin{equation}
\label{eq:kappa}
\epsilon \leq \kappa \Lambda_{QCD}^4 \Delta \bar\theta\big|_{\rm EDM} ,
\end{equation}
where $\kappa = 4\pi$ at the bottom of the potential and $\kappa = 4N$
in the linear regime. Substituting this equation into
Eq.~(\ref{eq:euc}), we see that the Euclidean action can be written as
\begin{equation}
S_0 = 5 \times 10^5 \left({f_a \over m_a}\right)^2 {1 \over
\kappa^3 (\Delta \bar \theta\big|_{\rm EDM})^3 }.
\end{equation}
The tunneling rate at the last stage is extremely suppressed for
parameters allowed by the constraints on the neutron EDM, $\bar
\theta|_{\rm EDM} <6\times 10^{-10}$.  To obtain a reasonable
tunneling rate, we need to get away from the thin wall limit (as
discussed previously); {\it i.e.} the value of $\epsilon$ must be
closer to $V_0 \sim (100 MeV)^4$ and hence must be larger than allowed
by Eq.~(\ref{eq:kappa}).  In order for Chain Inflation with the axion
to succeed, we must reconsider some assumptions we have made.

With large enough $N$, the tunneling rate can be
perfectly reasonable as long as one stays away from the absolute
bottom of the potential
[see Eq.~(\ref{eq:kappa}) with $\kappa=4N$ in the linear regime].  
Indeed, the
reheating of the universe can take place in the linear regime.  As
the
 field goes farther down the potential, the vacuum energy gets
smaller
 and smaller, and fewer $e$-folds result.  Hence, radiation
that is
 produced during reheating that takes place near, but not at,
the
 bottom of the potential, is not inflated away by the last few
episodes
 of tunneling to the bottom.  We have not yet investigated
details of
 the particles produced during reheating; axions may
provide the dark
 matter.

The only real problem with the
model is the last tunneling event, near the bottom of the potential.  In
Eq.~(\ref{eq:kappa}), with $\kappa=4\pi$ near the bottom,
we see that there is no dependence on $N$ in this regime.
The allowed energy difference is simply so small that no tunneling
takes place. The universe would still be situated in this false
vacuum now.

{\it Circumventing the Neutron EDM constraint:} The constraint on the
neutron EDM prevents the universe from tunneling in the last stage to
zero energy.  We may speculate about a number of ways to resolve this
problem. It is possible for the tunneling to stop at an energy of
$\epsilon_{bottom} \leq 10^{-3}$ eV and thereby account for the dark
energy. Alternatively, the soft breaking term and the QCD axion may
conspire to set $\gamma = 0$ in, avoiding the EDM constraint
altogether. Third, one might consider a time dependent tilt or a
different function for the tilt.  Alternatively, coupling to other
fields may allow the universe to tunnel or roll to the bottom via a
different direction in a two dimensional potential in field space.  Or
if the $Z_N$ symmetry breaks once the field is near the bottom of the
potential, then the field could quickly roll to zero energy.  Further
afield, axion models different from the DFSZ axion (see, e.g., the
review of \cite{kim}) might work better as inflaton candidates.
Briefly, there are models with several axions \cite{bgr,ck,knp} with
more desirable properties
\footnote{In superstring $E_8 \times E_8$
models, two axions are present.}.
Axions abound in string theory.  Some may solve the strong CP problem
\cite{banksdine1,banksdine2}. We have listed a few approaches
here to circumvent the neutron EDM constraints on tunneling to the bottom.

{\it On the Value of N.}
The value of $N$ need not equal the number of fermions that carry
color and CP charge.  The number of fermions may be far less,
depending on the relevant group representations. Defining
$T^{\alpha}_{(r)}$ to be the generators for the representation $r$,
one can write ${\rm Tr}(T^{\alpha\vphantom{\beta}}_{(r)}
T^{\beta}_{(r)}) = {1 \over 2} t_r \delta^{\alpha\beta}$.  The axion
model has an exact $Z_N$ symmetry with $N= (2 \pi/T_\Theta)
\bigl(\sum_{r,i} Q_{ri} t_r \bigr)$ where $Q_{ri}$ are the PQ charges
of the fermions and $T_\Theta$, the period of $\Theta$, need not be $2
\pi$ (see {\it e.g.} the review of Sikivie \cite{sirev}).

{\it Domain Walls.}
One might worry about the deleterious effect of domain walls which
appear when different horizon sized portions of the universe fall into
different minima at the QCD scale.  As shown by Sikivie
\cite{sikivie}, the energy difference between vacua leads to pressure
that causes the domain walls to move in the direction of eliminating
the higher energy vacua in favor of the lower energy ones.  The domain
walls disappear within a Hubble time. One might worry that the
universe is too quickly driven to zero energy, without any inflation,
but this does not happen.  By the Kibble mechanism, there are at most
two or three domains in any horizon volume, with different values of
$\langle a \rangle$.  Typically a horizon volume will be driven to a
field value half way down the potential, at $\sim N/2$, and will
subsequently inflate sufficiently.  In some part of the universe, a
horizon volume will contain only domains with values of $\langle a
\rangle \sim 0$, near the top of the potential.  Those regions that
start the highest up the potential will inflate the most.  Hence, if
one plunks down our observable universe in a random patch of the total
universe {\it after} inflation ends, then one is likely to find a
region that started inflating near the top of the potential and hence
inflated sufficiently.  Domain walls even have the positive effect of
driving a causal patch prior to inflation to become {\it more} uniform
by shoving away the nonuniformities.

{\it Conclusion.}
We have investigated using the QCD axion
potential to inflate.  We use the cosine shape of the axion, with $N$
minima, due to a residual $Z_N$ symmetry, together with a tilt
produced by a small soft breaking of the Peccei Quinn symmetry.  We
studied the DFSZ axion.  Chain inflation results along this tilted
cosine, with a series of tunneling events as the field tunnels its way
from a minimum near the top of the potential to ever lower minima.
Sufficient inflation as well as reheating result.  Tunneling in the
last stage is suppressed due to constraints on the neutron Electric
Dipole Moment and must be further considered.  We have 
listed a few attempts to work around these constraints.

In this paper, we have restricted discussion to axions which can solve
the strong CP problem. Obviously, if we forego any contact with real
QCD, then the allowed ranges for parameters becomes much larger.  For
example, the constraint from the neutron EDM vanishes. Then the ranges
of potential width, barrier height, and energy difference between
vacua are completely opened up.  A tilted cosine may arise due to
(non-QCD) ``axions'' in many other contexts, such as string theory,
and would easily provide an inflaton candidate.  
\smallskip

We thank A. Guth, H. Haber, G. Kane, P. Nilles, P.
Sikivie and E. Weinberg for useful conversations.  We thank the
Aspen Center for Physics as well as the Michigan Center for
Theoretical Physics for hospitality and support while this work was
completed.  This work was supported in part by the US Department of
Energy under grant DE-FG02-95ER40899.

\end{document}